# A new class of chiral materials hosting magnetic skyrmions beyond room temperature

Y. Tokunaga[1*], X.Z. Yu[1], J. S. White[2], H. M. Rønnow[3,1], D. Morikawa[1], Y. Taguchi[1] & Y. Tokura[1, 4]

[1] *RIKEN Center for Emergent Matter Science (CEMS), Wako, Saitama 351-0198, Japan,* [2] *Laboratory for Neutron Scattering and Imaging, Paul Scherrer Institut, CH-5232 Villigen, Switzerland,* [3]*Laboratory for Quantum Magnetism (LQM), École Polytechnique Fédérale de Lausanne (EPFL), CH-1015 Lausanne, Switzerland* [4]*Department of Applied Physics, University of Tokyo, Bunkyo-ku, Tokyo 113-8656, Japan * Present address: Department of Advanced Materials Science, University of Tokyo, Kashiwa 277-8561, Japan.*

**Abstract**

**Skyrmions, topologically protected vortex-like nanometric spin textures in magnets, have been attracting increasing attention for emergent electromagnetic responses and possible technological applications for spintronics. In particular, metallic magnets with chiral and cubic/tetragonal crystal structure may have high potential to host skyrmions that can be driven by low electrical current excitation. However, experimental observations of skyrmions have so far been limited to below room temperature for the metallic chiral magnets, specifically for the MnSi-type *B*20 compounds. Toward technological applications, it is crucial to transcend this limitation. Here we demonstrate the formation of skyrmions with unique spin helicity both at and above room temperature in a family of cubic chiral magnets: β-Mn-type Co-Zn-Mn alloys with a different chiral space group from that of *B*20**

**compounds. Lorentz transmission electron microscopy (LTEM), magnetization, and small angle neutron scattering (SANS) measurements unambiguously reveal the formation of a skyrmion crystal under the application of magnetic field ($H$<~1 kOe) in both thin-plate (thickness<150 nm) and bulk forms.**

Skyrmions with topologically non-trivial spin textures[1-3] have been observed or proposed to exist in various magnets due to different mechanisms, such as magnetic dipolar interaction[4,5], antisymmetric spin exchange or so-called Dzyaloshinskii-Moriya (DM) interaction[6,7,8,9], frustrated exchange interaction[10], and four-spin exchange interactions[11]. In metallic compounds, skyrmions induce fictitious magnetic fluxes acting on the conduction electrons, thereby inducing the so-called topological Hall effect[12-14]. Conversely, an applied electric current can induce a flow of skyrmions, with a threshold current density (~$10^6$ Am$^{-2}$) which is 5-6 orders of magnitude smaller than that necessary for driving magnetic domain walls in ferromagnets[15-17]. This property can be understood to arise from the theoretically predicted topological features of skyrmions[18-20]. On the other hand, in the insulating chiral magnet $Cu_2OSeO_3$, each skyrmion carries an electric dipole or quadrupole, which makes it possible to manipulate skyrmions by means of electric fields[21-25]. These features are promising for technological applications, including spintronic devices and ultra-low power-consumption, high-density magnetic memory.

Among the various skyrmions, those induced by DM interaction are particularly interesting due to their relatively small size (<150 nm) and fixed unique helicity (spin swirling direction). Theoretically, a number of helical magnets mediated by the DM interaction and with non-centrosymmetric crystallographic symmetries are predicted to be good candidates for hosting stable skyrmions[2,26]. In spite of such anticipation, experimental observations of skyrmions in non-centrosymmetric bulk magnets have been limited to only *B*20-type alloys[3,6,7,27] and $Cu_2OSeO_3$[21,22], all of which have the same cubic chiral space group $P2_13$. Their helical transition temperatures, and hence the maximum formation temperatures of skyrmions, are below room temperature (at most ~278 K in FeGe[27]).

$Co_{10}Zn_{10}$ with *β*-Mn-type structure[28,29], the target material system of the present study, belongs to another cubic chiral space group $P4_132$ or $P4_332$, depending on its handedness (see Figs. 1a and 1b). It contains 20 atoms in the unit cell and is composed of two kinds of crystallographic sites. One is 8*c* with the 3-fold site symmetry, and which is occupied by Co atoms, while the other is 12*d* with the 2-fold site symmetry and occupied by both Zn and Co atoms[29]. $Co_{10}Zn_{10}$ has been reported to be a ferromagnet with transition temperature ($T_c$) ~420 K and the Co magnetic moments (~0.85 $\mu_B$ per Co atom) ordered along <100> according to neutron diffraction measurements[29]. It has also been reported that the Co-Zn-Mn ternary system crystalizes in the *β*-Mn-type structure over a relatively wide composition range, and with $T_c$ dependent on the compositional ratio[30].

Similar to the case of *B*20 alloys[6,7,27] and $Cu_2OSeO_3$[21,22], the DM interaction combined with the chiral crystal structure might turn an otherwise ferromagnetic phase into a helimagnetic one with long magnetic periodicity. To confirm this and pursue the possibility of skyrmion formation under applied magnetic field, a series of several polycrystalline Co-Zn-Mn alloys were prepared and investigated.

The formation of chiral skyrmions is unambiguously demonstrated in both bulk and thin plate forms at and above room temperature in this metallic $\beta$-Mn-type Co-Zn-Mn alloy system, which can be an important step toward the technological application of skyrmions for spintronics devices.

**Results**

**Temperature dependence of magnetization.** For all prepared samples the structure was confirmed to be of $\beta$-Mn-type. Figure 1c shows the temperature dependence of the magnetization for selected samples measured in field-cooling runs. $T_c$ for $Co_{10}Zn_{10}$ is ~462 K, which is higher than the reported value (~420 K, ref. 29), possibly due to a slight difference in the composition. Upon substituting Co and Zn with Mn, $T_c$ systematically decreases and $Co_6Mn_6Zn_8$ no longer shows ferromagnetic-like behaviour. Magnetization measured at this magnetic field ($H$=20 Oe) is observed to decrease at low temperatures in many samples, but this tendency is mostly suppressed for $H$= 1 kOe. Whether there is a second transition or not at low temperature remain a topic for future investigations.

**Demonstration of the helimagnetic nature by SANS experiments.** To elucidate the possible helical nature of the magnetic ordering in these compounds, we have performed small angle neutron scattering (SANS) measurements. Figures 1d and 1e show typical SANS images obtained from polycrystalline pieces of $Co_{10}Zn_{10}$ and $Co_7Zn_7Mn_6$ under respective magnetic fields of 300 Oe and 250 Oe applied perpendicular to the incident neutron beam direction. In the vicinity of the $Q$=0 position, pairs of magnetic reflections are clearly observed nearly along the magnetic field direction, demonstrating the helical (or more precisely, conical due to the presence of the magnetic field) nature of the spin ordering in this system. The slight tilting of the peaks relative to the horizontal field direction demonstrates presence of weak magnetic anisotropy. As shown in Fig. 1f, the helical periodicity ($\lambda$) determined from the positions of the magnetic reflections shows a systematic change from $\lambda \sim 185$ nm for $Co_{10}Zn_{10}$ to $\lambda \sim 115$ nm for $Co_7Zn_7Mn_6$. This can be consistently understood as a decrease in the dominant ferromagnetic exchange interaction ($J$), which is reflected in the decrease of $T_c$. As a result, the periodicity which is given by $J/D$ (where $D$ is the DM interaction) also decreases. It is to be noted that all the compounds of $Co_{10-x/2}Zn_{10-x/2}Mn_x$ are good metals showing resistivity of 140-280 $\mu\Omega$cm at room temperature. A detailed transport study is currently in progress.

**Formation of the chiral skyrmion above room temperature.** Since the system is found to be a cubic DM helimagnet, the formation of skyrmions is anticipated upon application of the magnetic field in analogy to the case of $B$20-type (MnSi-type) alloys[6,7,27] and $Cu_2OSeO_3$[21,22]. To directly demonstrate the formation of skyrmions in this system, real-space observations by Lorentz transmission electron microscopy (LTEM) have been performed for a thin-plate sample with thickness of <150 nm. In LTEM the under- and over-focusing of the deflected electron beam can reveal magnetic contrast, which has proven an efficient way to investigate skyrmion textures[7]. A thin-plate specimen was sliced out from a polycrystalline ingot consisting of single-crystalline grains as large as several tens to hundred $\mu m^3$, and LTEM observation was performed for single-crystalline areas of larger than 100 $\mu m^2$ with specific crystallographic axis fortuitously normal to the thin-plate.

Figures 2a and 2b show the LTEM images for the (111) grain of $Co_8Zn_8Mn_4$ taken in the under-focus condition at 283 K. At $H=0$ (Fig. 2a), a magnetic stripe structure is observed with a periodicity ($\lambda \sim 124$ nm) that is in excellent agreement with the SANS result ($\lambda \sim 125$ nm). Upon applying $H=400$ Oe normal to the plate, emergence of a skyrmion crystal (SkX) is clearly observed.

Figures 2c-2f show the LTEM images obtained for a (110) grain of $Co_8Zn_{10}Mn_2$ at 345 K. Different from the former case of $Co_8Zn_8Mn_4$, there is no discernible magnetic contrast at $H=0$ in the region shown in Fig. 2c in spite of $T<T_c$, and magnetic stripes develop only in limited areas at this temperature (not shown). Considering the fact that LTEM is only susceptible to the in-plane components of the magnetic induction, which reflects in-plane magnetization, that does not cancel out when summed up along the thickness direction of the plate, a possible assignment of this phase is a helix with propagation vector directed normal to the plane, as conjectured from observations in the bulk B20 compounds[6,31]. Upon application of magnetic field, however, a SkX emerges despite the absence of a helical structure with in-plane propagation vector at $H=0$. Figures 2d and 2f are under- and over-focused images, respectively, taken at $H=650$ Oe, where the contrast is inverted from each other, confirming their magnetic origin. Figure 2e is a colour map of the in-plane magnetic induction (reflecting magnetization) components deduced from Figs. 2d and 2f by use of a transport-of-intensity equation (TIE) analysis[32]. Notably, the formation temperature (e.g. 345 K) of skyrmions for $Co_8Zn_{10}Mn_2$ here is well above room temperature.

Convergent-beam electron diffraction (CBED) patterns can be utilized to assign the crystal chirality of the region where the LTEM image of $Co_8Zn_{10}Mn_2$ was taken. Figure 2g shows the experimental CBED pattern with the [111] incidence, which has 3-fold symmetry. The simulated CBED patterns for the possible crystal chiralities of $P4_132$ and $P4_332$ are shown in Figs. 2h and 2i, respectively. The magnified images display significant differences that are dependent on the crystal chirality. The CBED pattern obtained experimentally (Fig. 2g) shows good agreement with the simulated pattern for the crystal chirality of $P4_132$ (Fig. 2h). The observed region was further shown to be a single crystallographic domain, and therefore of unique crystal chirality. This unique crystal chirality is hence reflected by the unique sign of the DM interaction so that the magnetic helicity of the skyrmions is fully aligned over the whole region (Fig. 2e), similarly to the case of the other cubic chiral magnets[7,21,27].

**Magnetic phase diagrams in the bulk and thin-plate form.** The signature of the SkX formation in bulk specimens was also observed by magnetization measurements. Figure 3a shows isothermal $dM/dH$ vs. $H$ curves measured at different temperatures for a polycrystalline piece of $Co_8Zn_9Mn_3$. Between ~311 K and ~320 K, a characteristic dip structure can be clearly seen at ~100 Oe, signalling the formation of the SkX in analogy with the $B$20-type alloys[33] and $Cu_2OSeO_3$ systems[21,22]. The magnetic phase diagram can be established from a contour plot of $dM/dH$, as shown in Fig. 3b. The SkX phase in the bulk appears in a relatively narrow temperature window with a width of ~9 K just below the helical ordering temperature. These features are in accord with the phase diagram of bulk $B$20 alloys[6] and $Cu_2OSeO_3$[21,22], and suggest that thermal fluctuations play a crucial role in stabilising the SkX phase also in this system of bulk form[6].

To observe directly the SkX and uncover the influence of dimensionality on its stability, the phase diagram was studied on a (111) thin plate (<150 nm) of $Co_8Zn_9Mn_3$ with use of LTEM. Figures 3c-3e and 3f-3h show typical LTEM images and their Fourier transformed images taken at $H$=0.7 kOe. At $T$= 295 K, a stripe structure with weak contrast in real space (Fig. 3c) and broad peaks in Fourier space (Fig. 3f) is observed, suggesting the system to be in the helical phase. At 315 K (Figs. 3d and 3g), the formation of the SkX with hexagonal symmetry is clearly observed. Further increase of the temperature to 320 K (Figs. 3e and 3h) results in the disappearance of magnetic contrast as the system enters the paramagnetic phase. Figure 3i shows a contour plot of the skyrmion density as a function of magnetic field vs. temperature. The magnetic field range to induce the SkX phase in the thin plate is significantly increased compared to the bulk case due to the demagnetization factor of the plate-like geometry[7,21,27]. Likewise, the SkX phase expands over a wider temperature range of ~20 K. As was the case for the other chiral magnets, this can be interpreted in terms of a relative suppression of the competing conical phase with out-of-plane propagation vector compared to the bulk case[7,21,27].

**SANS detection of the SkX formation in the bulk form.** To confirm the formation of the SkX in the bulk form more directly, SANS measurements were performed on a single-crystalline piece of $Co_8Zn_8Mn_4$ cut out from the polycrystalline ingot with fairly large grains obtained by the Bridgman method. Figure 4a shows the temperature dependence of the ac-susceptibility vs. magnetic field along the [110] direction. We conducted SANS measurements in the configuration where the directions of incident neutron beam and magnetic field were both approximately along the [110] direction. As shown in Fig. 4b, a ring-shaped intensity distribution can be seen at $T$=317 K and $H$=0, suggesting the formation of multi-domain single-$q$ helical structures with relatively flexible $q$-directions. At $H$=150 Oe, which is inside of the dip anomaly of ac-susceptibility (see Fig. 4a), a 6-fold-symmetric SANS pattern with magnetic reflections at $q\sim0.0058$ Å$^{-1}$ can be clearly observed (Fig. 4c), thereby directly demonstrating the formation of the triple-$q$ state or SkX with the lattice constant of

$$a_{Sk} = \frac{2}{\sqrt{3}}\lambda = \frac{4\pi}{\sqrt{3}q} \sim 125 \text{ nm}.$$

**Discussion**

The present observation of skyrmions in this $\beta$-Mn-type chiral magnet demonstrates the long-held expectation that new skyrmion hosting systems can be found in a variety of non-centrosymmetric crystal symmetries, and may as such stimulate experimental exploration of other realizations, including other $\beta$-Mn structured materials. Likewise, the long sought-after demonstration of skyrmion stabilisation above room temperature implies that complicated cooling is no longer a limiting factor for the potential integration of skyrmions into technological spintronics devices and applications.

**Methods**

**Sample preparation.** Polycrystalline samples of Co$_x$Zn$_y$Mn$_z$ ($x+y+z=20$) were synthesized by the method described in ref. 29. Stoichiometric amounts of pure Co, Zn, and Mn pieces of 2 g total mass were sealed in evacuated quartz tubes ($\sim 10^{-3}$ Pa), heated to 1000 °C for 12h, followed by slow cooling (1 °C/h) down to 925 °C, then kept at this temperature for 75~100 h. Finally, the products were water quenched. A polycrystalline sample of $Co_8Zn_8Mn_4$ with larger grain sizes (typically up to 3 mm) was grown by the Bridgman method with cooling from 1025 °C to ~700 °C over one week, followed by water quench. A single-crystalline specimen was cut out from the resultant product and oriented with use of back-reflection x-ray Laue photographs.

**Magnetization measurements.** Magnetization and ac-susceptibility were measured by a superconducting quantum interference device magnetometer (MPMS3, Quantum Design) equipped with oven and ac-susceptibility measurement options.

**Small-angle neutron scattering (SANS) measurements.** The long wavelength magnetic structures in bulk samples of $Co_xZn_yMn_z$ were measured using the small-angle neutron scattering (SANS) instrument SANS-I at the Swiss Spallation Neutron Source (SINQ), Paul Scherrer Institut, Switzerland. In a typical instrument setup, neutrons of wavelength 10 Å were selected with a FWHM spread ($\Delta\lambda/\lambda$) of ~10%, and collimated over a distance of 18 m before the sample. The scattered neutrons were collected by a position-sensitive multi-detector placed 18 m behind the sample. Polycrystalline ingot samples of $Co_xZn_yMn_z$, and single crystals samples of $Co_8Zn_8Mn_4$, were mounted inside a horizontal field cryomagnet that was installed on the beamline. The cryomagnet could be rotated so that the direction of applied magnetic field was either approximately parallel, or approximately perpendicular to the neutron beam. For SANS measurements on the polycrystalline samples, the magnetic field was applied perpendicular to the neutron beam (Figs. 1c and 1d), while measurements on the single crystal samples made use of an applied field approximately parallel to the neutron beam (Figs. 4b and 4c).

For either experimental geometry, the SANS measurements were done typically by rotating the sample and cryomagnet ensemble in a step-wise manner around its vertical axis - a so-called rocking scan - and over a sufficiently broad angular range so as to move the various magnetic diffraction spots through the Bragg condition at the detector. Typical rocking scans spanned an angular range of up to ±20° with a SANS measurement conducted every 2º. This chosen step size was larger than the calculated instrumental contribution to the rocking half-width of the magnetic Bragg peaks (~0.2º), yet smaller than the observed Lorentz factor-corrected rocking half-widths which, while sample dependent, were always greater than ~4.5º. This observation demonstrates the observed rocking widths of the magnetic peaks to be dominated by the mosaicity of the magnetic order in the sample. By summing the detector measurements taken over all rotation angles, all of the diffraction spots can be seen in a single image, thus resulting in the SANS diffraction patterns shown in Figs. 1 and 4.

**Determination of the crystallographic chirality.** The crystal chirality of $Co_8Zn_{10}Mn_2$ was determined by using convergent-beam electron diffraction (CBED) pattern. The CBED pattern was obtained by using a JEM-2100F at an accelerating voltage of 200 kV at room temperature. For the simulation of CBED patterns, the software MBFIT[34] was used. From the comparison between the experimental and simulated CBED patterns, the crystal chirality can be determined uniquely.

**Lorentz transmission electron microscopy (LTEM) measurements.** For Lorentz transmission electron microscopy (LTEM) measurements, thin plates were cut from polycrystalline samples and thinned by mechanical polishing and argon ion milling. The crystal orientation and thickness were confirmed by conventional TEM observation and the electron energy loss spectroscopy (EELS), respectively. In LTEM observations, the magnetic structures can be imaged as convergences (bright contrast) or divergences (dark contrast) of the electron beam on the defocused (under- or over-focused) image planes. The inversion of contrast between the over- and under-focused images as well as disappearance of the contrast above $T_c$ assured the magnetic origin of the image contrast. The analysis of the in-plane components of magnetic induction (which reflects the in-plane magnetization) from the LTEM data was performed by the transport-of-intensity equation (TIE) method with use of the software package QPt (HREM Co.)[32].

**Acknowledgements**

The authors thank A. Kikkawa for help in sample preparation. This work was supported by JSPS Grant-in-Aids for Scientific Research(S) No.24224009 and for Young Scientists (A) (No. 25707032). Neutron scattering was performed at the Swiss Spallation Neutron Source (SINQ). Work in Switzerland was supported by the Swiss National Science foundation and the European Research Council project CONQUEST. HMR gratefully thanks CEMS for the hospitality during this work.

**Author contributions**

The sample preparation and magnetic characterization were performed by Y. Tokunaga. LTEM observations were performed by X.Y. CBED patterns were obtained and analyzed by D.M. Small angle neutron scattering measurements were conducted by J.W, H.M.R., and Y. Tokunaga. The results were discussed and interpreted by all the authors.

**Additional information**

Correspondence should be addressed to Y. Tokunaga. (e-mail: *y-tokunaga@riken.jp*).

**Figure legends**

Figure 1: **Crystal structure, magnetic properties, and helical magnetic order in the $\beta$-Mn-type Co-Zn-Mn alloys.** (**a,b**) Schematic of $\beta$-Mn-type structure with (**a**) $P4_132$ and (**b**) $P4_332$ space group, dependent on its handedness (chirality). (**c**) Temperature dependence of the magnetization for a series of polycrystalline Co-Zn-Mn alloys measured at $H$=20 Oe upon cooling. (**d,e**) Small angle neutron scattering (SANS) images for polycrystalline samples of $Co_{10}Zn_{10}$ and $Co_7Zn_7Mn_6$ at $T$=317 K, $H$=300 Oe and $T$=120 K, $H$=250 Oe, respectively. The magnetic field direction is perpendicular to the neutron beam, which is itself into the page. (**f**), Composition dependence of the helical ordering temperature

(filled blue circles) and helical pitch obtained by the SANS measurements (filled red squares) and LTEM observations (open red circles), respectively.

Figure 2: **Lorentz transmission electron microscopy (LTEM) observation of skyrmion crystal (SkX) formation by application of a magnetic field.** (**a**, **b**) LTEM image at 283 K on (111) plane of $Co_8Zn_8Mn_4$ taken with under-focused condition at $H$=0 and 400 Oe, respectively. (**c, d, f**) LTEM image on (110) plane of $Co_8Zn_{10}Mn_2$ taken with under-focused condition obtained at (**c**) $H$=0 and (**d**) 650 Oe, and (**f**) with over-focused condition at $H$=650 Oe. In (c), the dark band running diagonally across the image is the diffraction contrast and not of magnetic origin. (**e**). Colour map of in-plane components of the magnetic induction (which reflects in-plane magnetization) at $H$=650 Oe deduced from TIE analysis of the images (**d**) and (**f**). (**g**) Observed convergent-beam electron diffraction (CBED) pattern of $Co_8Zn_{10}Mn_2$ taken with the [111] incidence from the same specimen area where the LTEM images (**c-f**) were obtained. (**h, i**) Simulated CBED patterns assuming the space group No. 213 ($P4_132$) and No. 212 ($P4_332$), respectively. Below each panel (**g-i**), magnified images for the selected area are shown. The central one of the magnified images of g is based on another picture with different exposure time for the purpose of avoiding intensity saturation. Magnified images of the simulated pattern show significant differences for each crystal chirality. The simulated CBED pattern using $P4_132$ shows a good agreement with the experimental one.

Figure 3: **Comparison between magnetic phase diagrams of bulk and thin plate forms of $Co_8Zn_9Mn_3$.** (**a**) Temperature dependence of isothermal $dM/dH$ vs. $H$

curves for a polycrystalline piece of $Co_8Zn_9Mn_3$. (**b**) Magnetic phase diagram of the bulk sample in magnetic field vs. temperature plane as deduced from *dM/dH* curves. (**c,d,e**) LTEM images on (111) plane of $Co_8Zn_9Mn_3$ taken with under-focused condition in *H*=0.7 kOe applied normal to the plane at (**c**) *T*=295 K, (**d**) 315 K, and (e) 320 K, respectively. (**f,g,h**) Diffractographs (Fourier transforms) of the LTEM images (**f**) **c**, (**g**) **d** and (**h**) **e**, respectively. (**i**) Contour plot of skyrmion density of the $Co_8Zn_9Mn_3$ (111)-thin plate in *H* vs. *T* plane as deduced from LTEM observations obtained by counting the number of skyrmions in the area of ~120 $\mu m^2$ at each temperature and magnetic field. Some regions within the area contained dense SkX like in image (**d**) but other regions did not, making the averaged density smaller than expected from the image (**d**). In the present bulk magnetization measurement, demagnetization factor (*N*) is ~0.1 from the sample shape, whereas *N*~1 in the thin plate form for LTEM observation. Taking the magnitude of the magnetization in this sample into consideration, it explains factor of 7~8 difference in the threshold external magnetic field for formation of the SkX phase in the bulk form (**b**) and thin-plate form (**i**).

Figure 4: **Demonstration of the formation of the SkX in the bulk form by means of small angle neutron scattering (SANS) measurements.** (**a**) Temperature dependence of ac-magnetic-susceptibility vs. magnetic field measured in *H*||[110] for a single-crystalline piece of $Co_8Zn_8Mn_4$. (**b, c**) SANS images at 317 K obtained at *H*=0 and *H*=150 Oe, respectively. *H* and the incident neutron beam are parallel with each other and both along [110] direction. The horizontal and vertical axes of the SANS images correspond to the [001] and [1-10] crystal directions, respectively.

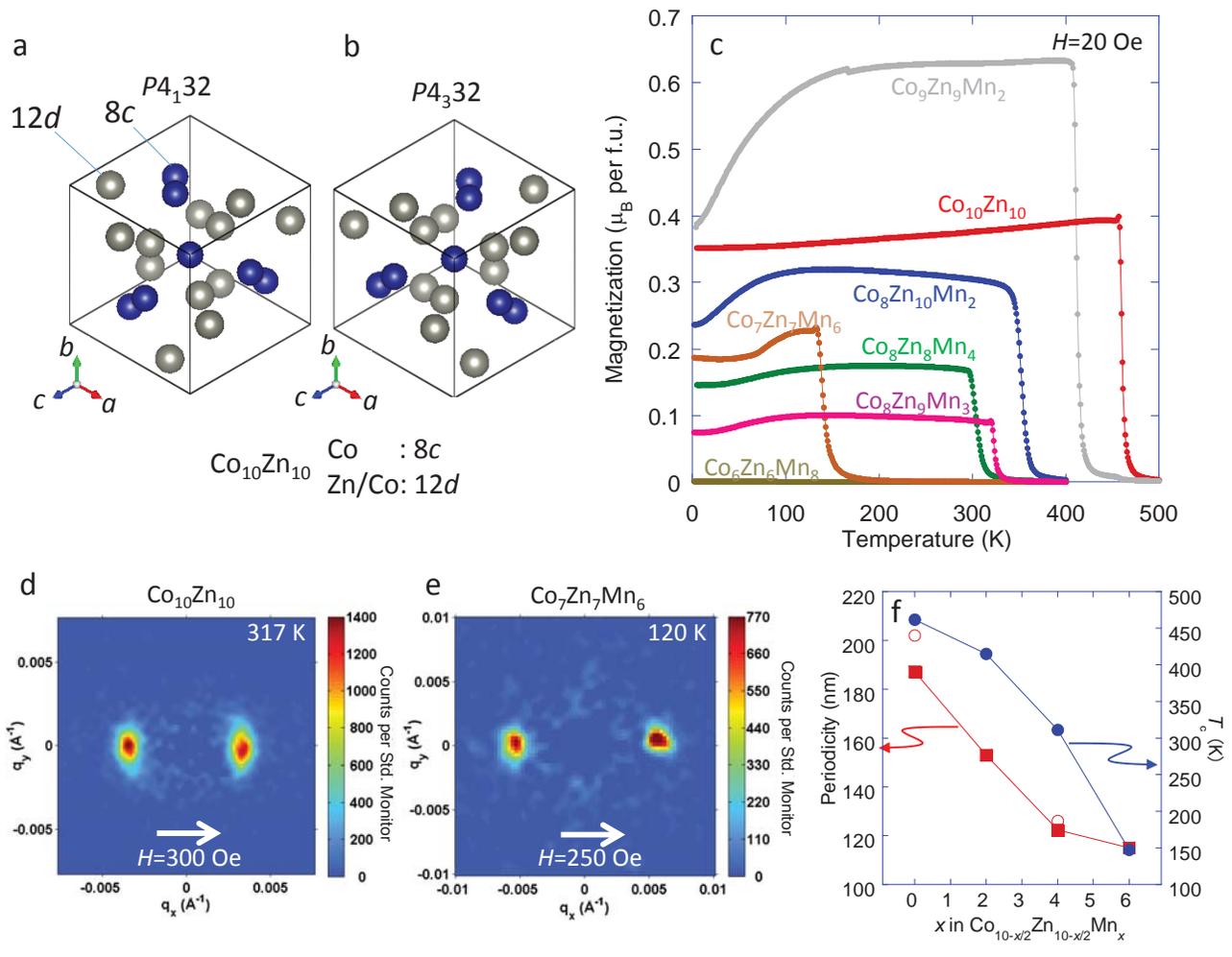

Fig. 1

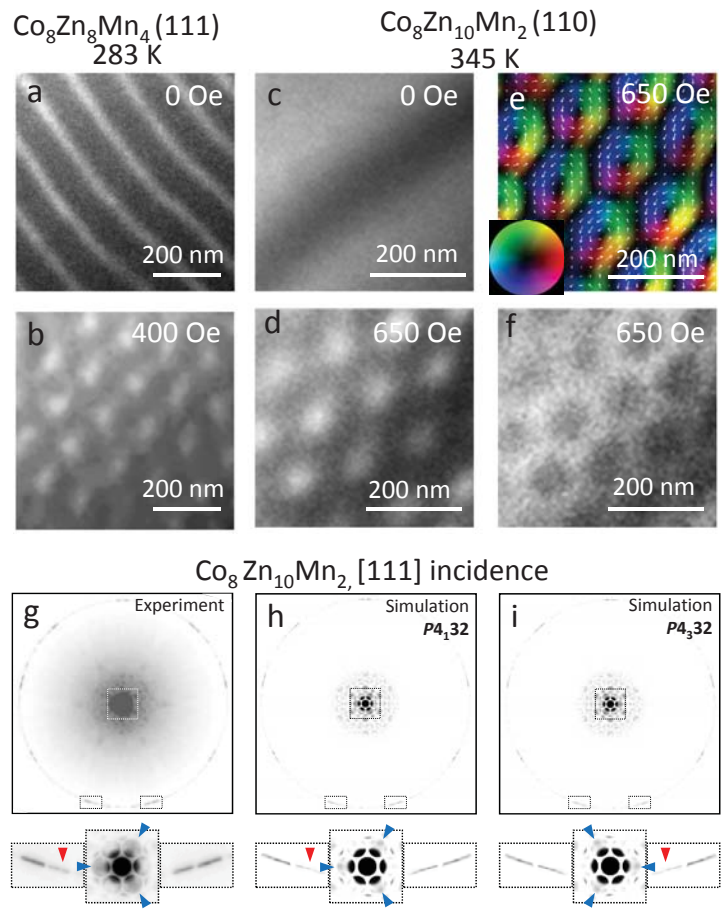

Fig. 2

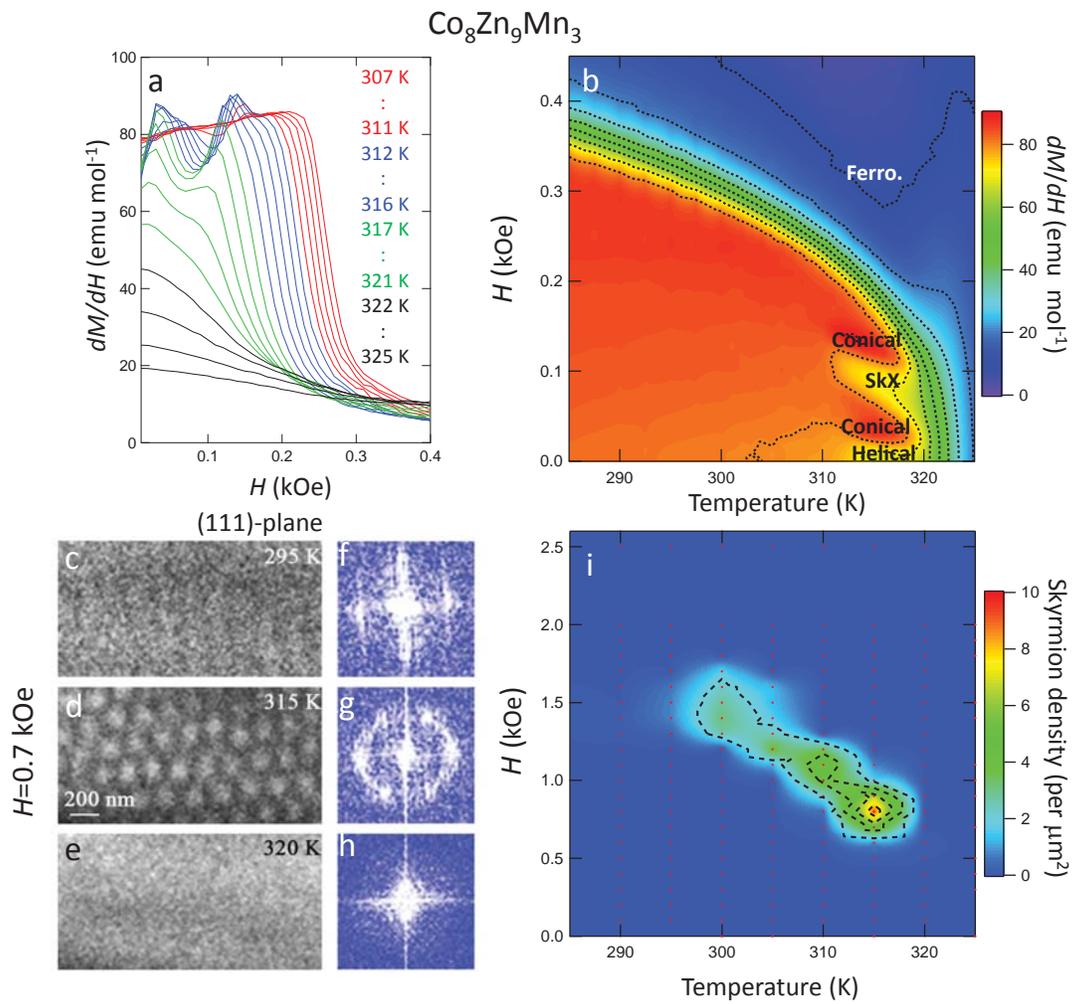

Fig. 3

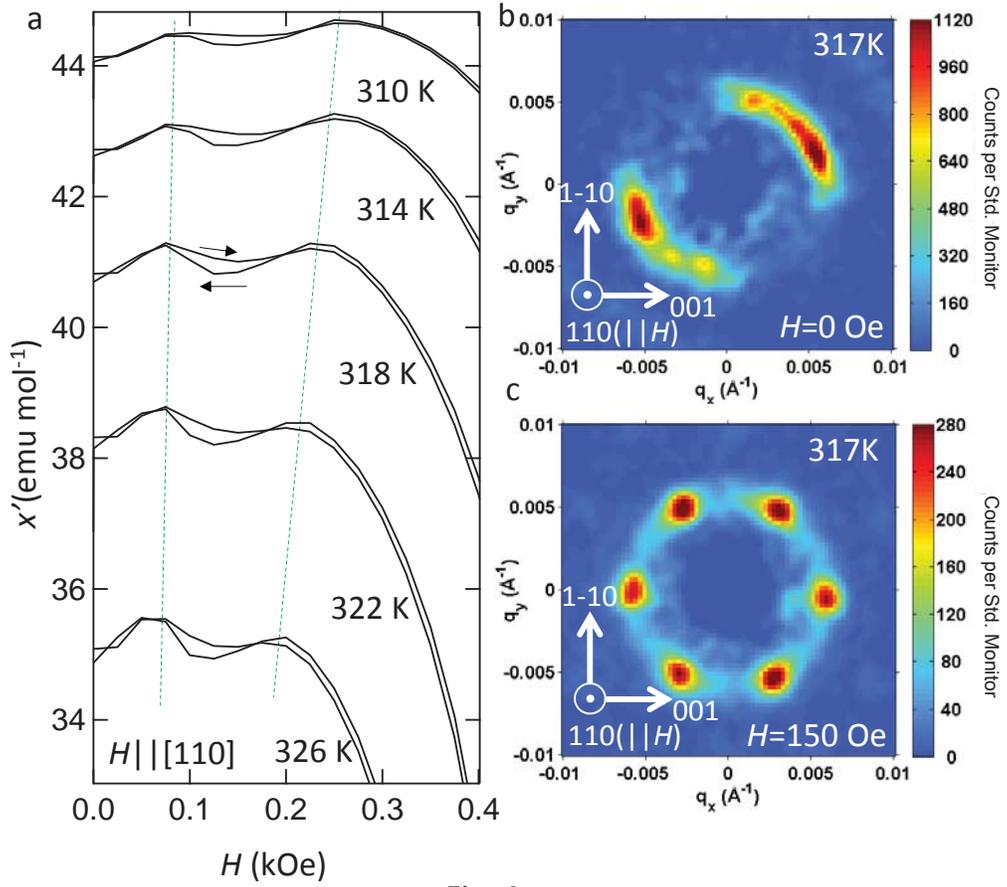

Fig. 4